# Scalable fabrication of single silicon vacancy defect arrays in silicon carbide using focused ion beam

Junfeng Wang,[1] Xiaoming Zhang,[1] Yu Zhou,[1] Ke Li,[1] Ziyu Wang,[1] Phani Peddibhotla,[1] Fucai Liu,[2] Sven Bauerdick[3] and Axel Rudzinski,[3] Zheng Liu,[2] and Weibo Gao[1,4,5*]

[1]*Division of Physics and Applied Physics, School of Physical and Mathematical Sciences, Nanyang Technological University, Singapore 637371, Singapore*
[2]*Center for Programmable Materials, School of Materials Science & Engineering, Nanyang Technological University, 50 Nanyang avenue, Singapore 639798, Singapore*
[3]*Raith GmbH, 44263 Dortmund, Germany*
[4]*The Photonics Institute and Centre for Disruptive Photonic Technologies, Nanyang Technological University, 637371 Singapore, Singapore*
[5]*MajuLab, CNRS-University of Nice-NUS-NTU International Joint Research Unit UMI 3654, Singapore*
*\*Corresponding author: wbgao@ntu.edu.sg*

**Abstract**

In this work, we present a method for targeted and maskless fabrication of single silicon vacancy ($V_{Si}$) defect arrays in silicon carbide (SiC) using focused ion beam. Firstly, we studied the photoluminescence (PL) spectrum and optically detected magnetic resonance (ODMR) of the generated defect spin ensemble, confirming that the synthesized centers were in the desired defect state. Then we investigated the fluorescence properties of single $V_{Si}$ defects and our measurements indicate the presence of a photostable single photon source. Finally, we find that the $Si^{++}$ ion to $V_{Si}$ defect conversion yield increases as the implanted dose decreases. The reliable production of $V_{Si}$ defects in silicon carbide could pave the way for its applications in quantum photonics and quantum information processing. The resolution of implanted $V_{Si}$ defects is limited to a few tens of nanometers, defined by the diameter of the ion beam.

Silicon carbide (SiC) is a technologically mature semiconductor material, which can be grown as inch-scale high-quality single crystal wafers and has been widely used in microelectronics systems and high-power electronics, etc. In recent years, some defects in SiC have been successfully implemented as solid state quantum bit[1-8] and quantum photonics [9-11]. They meet essential requirements for spin-based quantum information processing such as optical initialization, readout and microwave control of the spin state, which are similar as the nitrogen vacancy (NV) centers in diamond. [12] In particular, silicon vacancy ($V_{Si}$) defect in 4H-SiC has increasingly attracted attention owing to its excellent features, such as non-blinking single photon emission and long spin coherence times which persist up to room temperature (about 160 μs). [3,5,13] These remarkable properties have been exploited in many applications in quantum photonics, [9,10] and quantum metrological studies such as high sensitivity magnetic sensing[14,15] and temperature sensing.[16]

The $V_{Si}$ defect consists of a vacancy on a silicon site which exhibits a $C_{3v}$ symmetry in 4H-SiC.[3,5] In order to extend its applications in quantum information science, it is essential to develop the technique of scalable efficient generation of single $V_{Si}$ defect arays in 4H-SiC. Since the collected fluorescence rate of a single $V_{Si}$ defect is modest with only about 10 kcps,[3,5,17] it is required to couple with some photonic devices to improve the counts towards the construction of photonics networks.[3,9,10,17,20] However, in order to realize the mode-maximum of photonic devices, it is necessary to place the $V_{Si}$ defects relative to the optimal position with sub-wavelength-scale precision. Previously there are three methods to generate $V_{Si}$ defect: the electron irradiation, neutron irradiation, and carbon implantation, however, these methods either can't control the position of the $V_{Si}$ defect, or need a electron beam lithography (EBL) pre-fabricated photoresist patterned mask, [3,5,9,17] which is not convenient for coupling to pre-fabricated photonic devices.

To solve this issue, in this work, we report a new approach to fabricate maskless and targeted single $V_{Si}$ defects array in 4H-SiC using silicon focus ions beam implanted (FIB). This scalable post-fabrication approach can control the $V_{Si}$ defect position accuracy in tens of nanometers, which can greatly simplify the design and fabrication process and improve the efficiency of $V_{Si}$ defects based photonics structure fabrication.[9,20] We used different dose (ions/spot) to create $V_{Si}$ defect arrays and stripes (which serve as markers for reference) using 35 keV focused $Si^{++}$ ions beam implantation. Firstly, we studied the room and cryogenic temperature photoluminescence (PL) spectrum and optically detected magnetic resonance (ODMR) of the ensemble defects on the stripe to confirm that they are $V_{Si}$ defects. Then we studied the optical properties of a single $V_{Si}$ defect in the lowest dose (40 Si/spot) area, and found it is a single photon source with good photo-stability. Moreover, we systematically investigate the mean counts per spot and the $Si^{++}$ ion to $V_{Si}$ defect conversion yield as a function of the implanted dose. Our results demonstrate that this method is well adapted to the targeted scalable generation of $V_{Si}$ defects, paving ways towards $V_{Si}$ defect based scalable quantum photonic devices.

A commercial 40 keV nanoFIB system (ionLINE Plus, RAITH Nanofabrication) was used for the focused Si ion beam implantation. The nanoFIB system creates a tightly focused ion beam with a diameter of a few nanometers from liquid metal alloy ion sources, which ensure the implanted Si ions lateral distribution at nanoscale. Moreover, it can also realize parallel Si ions implantation about tens of thousands of sites per second, which allows scalable fabrication of $V_{Si}$ defects in the sample.[20] A commercial high-purity 4H-SiC sample with a low background fluorescence was used for the experiments.[3,5,17] A 35keV double charged silicon ions ($Si^{++}$) were implanted onto the sample to generation $V_{Si}$ defect array. The implantation spots were targeted with between 40 to 700 Si ions with a 5-10nm nanometers focus ion beam spot. [19-21] Two stripes with their width about 1.5 μm was also generated by 700 Si ions/ spot (20nm step size in each direction), which were used for marking. The ion current was controlled by both the beam current and the implantation time. After annealing the implanted defect array sample at 650 ℃ in air for 6 hours, we put the sample in $HNO_3$ for 24 hours to remove amorphous SiC layer in order to decrease the fluorescence background.[3] The positioning accuracy of the $V_{Si}$ defect is determined by the diameter of the focused ion beam (5-10nm) and the lateral straggling effects of implanted silicon ions in SiC. Inferred from SRIM simulations, the average depth of the silicon ions was 18.5nm, and its longitudinal and lateral straggling uncertainty was about 7 and 6 nm respectively. So it is estimated that the spatial positioning precision of the $V_{Si}$ defect is less than 100nm in all three dimensions. [19-22]

Firstly, we studied the optical and spin properties of the focused ion beam implanted $V_{Si}$ defects using a home-built confocal fluorescence microscopy system.[17] We used a 740nm laser for optical pumping of these defects as the wavelength of this laser is close to the optimum excitation wavelength of 770 nm.[23] We also used high quantum efficiency single photon detectors (ID Quantique ID120) in order to enhance the fluorescence collection of the $V_{Si}$ defects. Fig. 1(a) shows the PL scan of a representative area of the 700 Si ions implanted $V_{Si}$ defects array and stripes on the SiC surface (laser power 0.7mW). The spots are separated by ~2 μm. In order to identify the type of the defects, we firstly measured the room temperature PL spectrum of an ensemble of defects on the stripe after filtering the emission through a 900nm longpass filter (Fig1. (b)). The observed spectrum is in the near-infrared range and also similar to the previous measurement results of the PL spectrum of single and ensemble $V_{Si}$ defects.[3,5,7,17] Further, we measured the cryogenic temperature (5K) PL of the ensemble defects on the stripe with a cryogenic temperature confocal system (Montana Instruments Cryostation). [17,24] As shown in Fig. 1(c), there are two spectral emission peaks (861.4nm, 916.5nm) which correspond to the two inequivalent lattice sites of the $V_{Si}$ defects in 4H-SiC: V1 (861.4 nm) and V2 (916.3 nm) centers respectively. [7,17,24]

The negatively charged V2 center in 4H-SiC is $V_{Si}$ defect with spin 3/2. For spin manipulation of the implanted $V_{Si}$ defects, we measured the continuous-wave ODMR of the ensemble $V_{Si}$ defects on the stripe at room temperature. The radio frequency signal generated by a microwave source was first gated though a switch (ZASWA-2-50DR), and then amplified by a high power amplifier (ZHL-20W-13+). The radio frequency irradiation of $V_{Si}$ defects was achieved with a 20μm width microwave stripline fabricated on the surface of sample by standard lithography techniques.[17] The experiment was synchronized by a pulse generator (PulseBlaster PBESR-PRO-500). Since the ODMR contrast is expected to be less than one percent, we measured the ODMR signal by modulating the microwave drive amplitude (ON, OFF) which had a duty cycle of 0.5 and a half cycle duration of 2.8ms.[7,9,17] Finally, the ODMR contrast, related to the change in PL, is calculated using the formula ΔPL = (ΣN(ON)- ΣN(OFF))/ ΣN(OFF).[9,17] Fig 1(d) shows the results of the ODMR measurements of the $V_{Si}$ ensemble defects and the red curve is a Lorentzian fit to the data. From the fit, we obtain an estimated resonant frequency of 70.2 MHz, which is close to the zero field splitting of the $V_{Si}$ center (2D = 70 MHz).

Next, we studied the single $V_{Si}$ defect properties in the lowest dose (40 Si ions) implanted area. Fig. 2(a) shows the PL image of the $V_{Si}$ defects array obtained from the silicon-implanted region at the SiC surface (laser power 0.7mW). The silicon ion dose was 40 ions per spot and the distance between spots was 2 μm. In order to identify a single $V_{Si}$ defect, we measured the second-order autocorrelation function $g^2(\tau)$ using the Hanburry-Brown-Twiss (HBT) setup.[17] Since the fluorescence count rate is low (13kcps), the effect of the background fluorescence has to be included in the evaluation of $g^2(\tau)$. The auto-correlation function $g^2(t)$ is corrected from normalized raw data $C_N$ (t) using the function $g^2(\tau) = (C_N(\tau)-(1-\rho^2))/\rho^2$ , where $\rho$ = s/(s+b), and s and b are the signal and background count, respectively.[5,9,17] In the experiment s and b was about 3 kcps and 2 kcps, respectively. A typical background-corrected $g^2(\tau)$ was shown in Fig. 2(b), and the red line was the fit according to the function $g^2(\tau) = 1-a*exp(-|\tau|/\tau_1)+ b*exp(-|\tau|/\tau_2)$, where a, b, $\tau_1$, $\tau_2$ are fitting parameters. Form the fit, we got $g^2(0)$ = 0.35 ± 0.05, which demonstrated it was a single $V_{Si}$ defect.

The property of photostability is important for single photon sources in various quantum information applications. [3,5] To this end, as shown in Fig 2(c), we measured the fluorescence intensity trace of a single $V_{Si}$ defect with a sampling time of 100ms and 1mW laser power for a duration of 110 s. We could see that its count was very stable, and had no blinking or bleaching, which verified that it is a stable single photon source. Fig. 2(d) shows the emission intensity saturation curve of a single $V_{Si}$ defect. The red line is a fit according to the formula $I(P) = I_s/(1+P_0/P)$, where $I_s$ is the maximum photon count rate, and $P_0$ is the saturation power, P is the laser power. Inferred from the fit, the maximum count rate $I_s$ was about 13.0 kcps, and saturation power $P_0$ was about 0.48mW.

Efficient generation of defects in SiC is important for the coupling with photonics structure,[9]. In view of this, we present statistics of the distribution of the number of $V_{Si}$ defects for 50 implanted spots by measuring defects $g^2(0)$ values, PL intensity and the mean counts of a single $V_{Si}$ defect. The results are shown in Fig. 2(e), which agree with a Poisson distribution, $P_\lambda(k) = (\lambda^k/k!)/e^{-\lambda}$. Herr λ is the average number of $V_{Si}$ defects per spot and k is the number of $V_{Si}$ centers actually formed in the spot. The average number of $V_{Si}$ defects per spot, λ, was evaluated to be 1.56 by fitting the obtained distribution with a Poisson distribution. Thus the conversion yield of $V_{Si}$ defects formation is evaluated to be about 3.9 %. The conversion yield is comparable with the focused ion beam to generate the SiV center in diamond. [20,22] Moreover, the number of the single $V_{Si}$ defect was about 19, so the single $V_{Si}$ defect generation efficiency was about 38 %.

Finally, in order to systematically investigate the conversion yield of implanted $Si^{++}$ ions to $V_{Si}$ defects, we varied the ion doses on the sample from 40-700 $cm^{-2}$. The ion dose determined the number of silicon vacancies created during the implantation process. We measured the fluorescence intensity across a 10×10μm² region of constant implantation dose, and normalized to the mean counts per spot. Fig. 3(a) and (b) show typical confocal fluorescence image of an area of the 100 and 400 Si ions implanted $V_{Si}$ defects array on the SiC surface, respectively. We observe that the PL emission intensity associated with $V_{Si}$ centers increases with increasing silicon ion dose indicating an increase in the number of $V_{Si}$ defects produced. As shown in Fig. 3(c), the average fluorescence rates per spot increases with increasing implantation dose (ions/spot). We found that when the implanted dose changes from 40 to 100 ions/spot, the mean counts per spot increased quickly. Then, it increased slowly with the implanted dose from 100 to 700 ions/spot. Finally, we normalized the mean counts per spot to the corresponding implanted ion number to get the conversion yield. As shown in Fig. 3(d), the conversion yield was almost same when the dose was 40 to 100 ions/spot, then it decreased as the dose increased from 100 to 700 ions/spot. This reduced yield might result from the accumulation of lattice damage and charged defects in SiC lattice, similar to what was observed in SiV centers in diamond.[20]

In conclusion, we present a method for scalable and maskless generation of arrays of silicon vacancy centers in SiC using focused $Si^{++}$ ion beam technology. The measured PL spectrum and ODMR of defect spin ensembles prove that the created defect centres are mostly the $V_{Si}$ defect spins. Then we investigated the fluorescence properties of a single $V_{Si}$ defect and found that it is a photostable single photon source with a saturation count rate of up to 13kcps. Finally, we found that the conversion yield of $Si^{++}$ ions to $V_{Si}$ defects decreases for higher implantation doses. This method offers new possibilities for realizing photonics applications where there arises a need to place single or a few $V_{Si}$ defect centers within a nanometre-scale region. The photonic structures that may take benefit of this new method are solid immersion lenses (SIL),[3] nanopillars[9] or photonic crystal structures,[10] fibers,[22] as well as waveguides.[18,25] Single $V_{Si}$ defects could also be incorporated into the intrinsic region of a SiC positive-intrinsic-negative (PIN) diode structure for single-photon emission[26] and also into opto-mechanical resonators.[27] It can also be used to generate dipolar coupled $V_{Si}$ defect pairs facilitating the creation of quantum registers.[28,29]

**Acknowledgment.** We thank Evan Toh and Andrew Yu for the discussion about nanofabrication. We acknowledge the support from Singapore National Research Foundation through a Singapore 2015 NRF fellowship grant (NRF-NRFF2015-03, NRF-RF2013-08), its Competitive Research Program (CRP Award No. NRF-CRP14-2014-02), A*STAR QTE Project and Singapore Ministry of Education tier 1 project RG176/15.

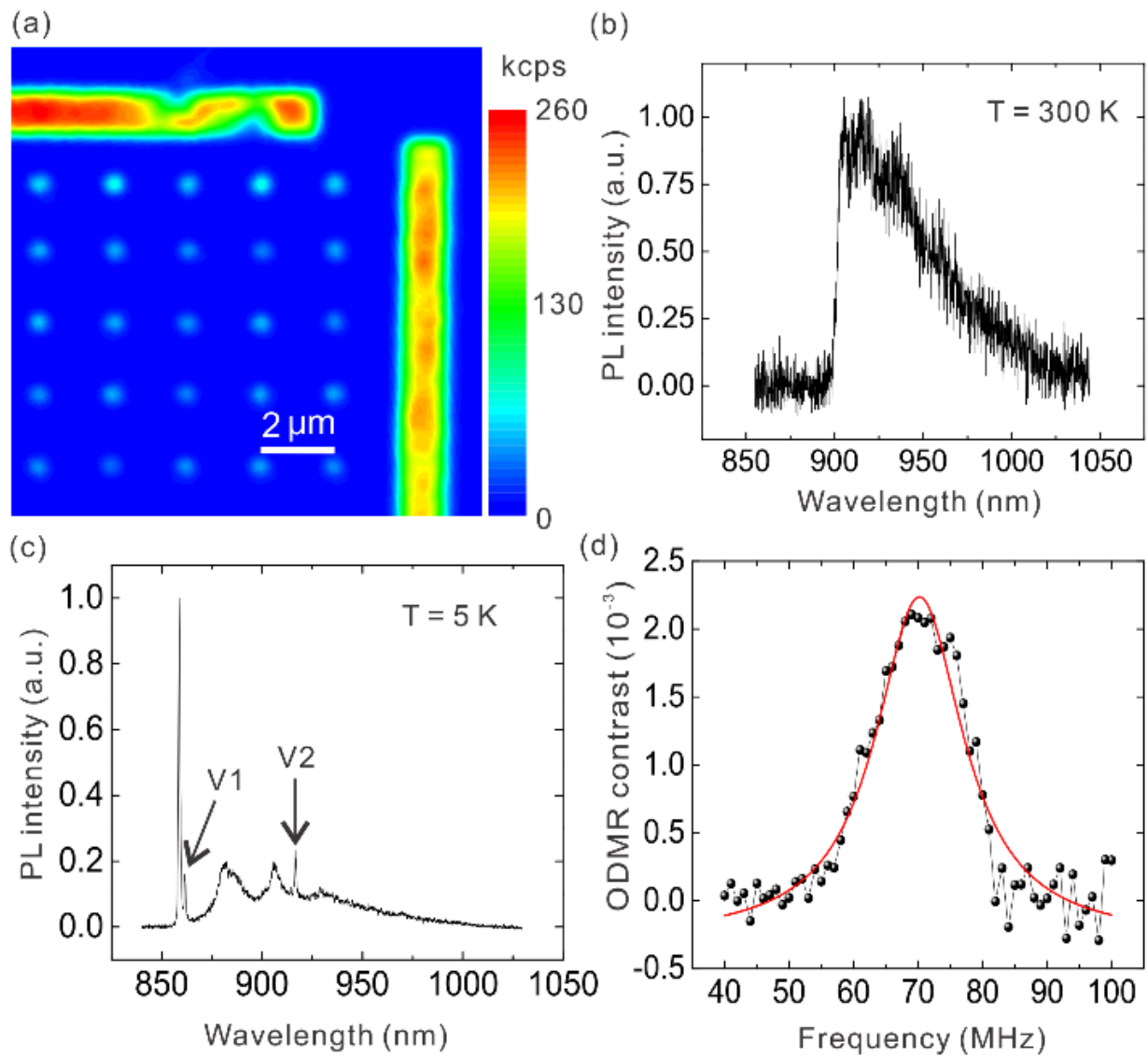

Figure1. Characterization of ensembles of defect. (a) Confocal fluorescence image for an area of the 700 Si ions implanted $V_{Si}$ defects array and stripe on the SiC surface. The scale bar is 2μm. (b) Room temperature PL spectrum measurement of the $V_{Si}$ defects on the stripe. (c) The PL spectrum of the $V_{Si}$ defects on the stripe at cryogenic temperature (5K). The two peaks correspond to the zero phonon lines of the V1 and V2 centers of the $V_{Si}$ defects. (d) The room temperature ODMR detection of $V_{Si}$ defects. The red line is the fit of the data.

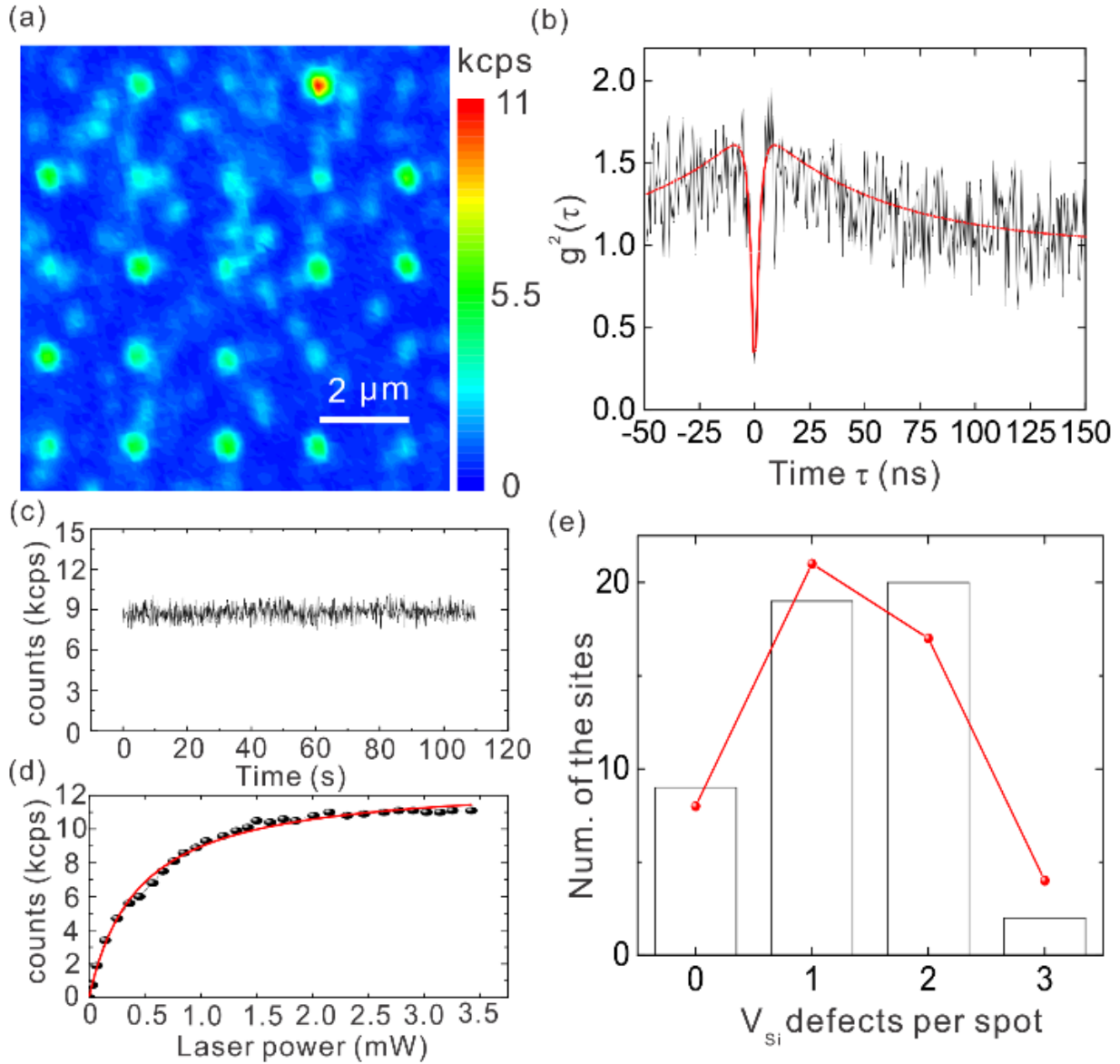

Figure 2. Single defect characterization. (a) Confocal fluorescence image for an area of the 40 Si ions implanted single $V_{Si}$ defects array on the SiC surface. The scale bar is 2μm. (b) Background-corrected second-order autocorrelation function of a single $V_{Si}$ defect with $g^2(0) = 0.35 \pm 0.05$ under 0.65 mW of laser excitation power. (c) The fluorescence intensity trace of the single $V_{Si}$ defect with a sampling time of 100ms and 1mW laser excitation power. (d) The fluorescence intensity measurements of a single $V_{Si}$ defect emitter at different laser powers. (e) Statistics of the number of $V_{Si}$ defects per implanted spot evaluated on 50 implanted sites. The data is fitted by a Poisson distribution (red curve).

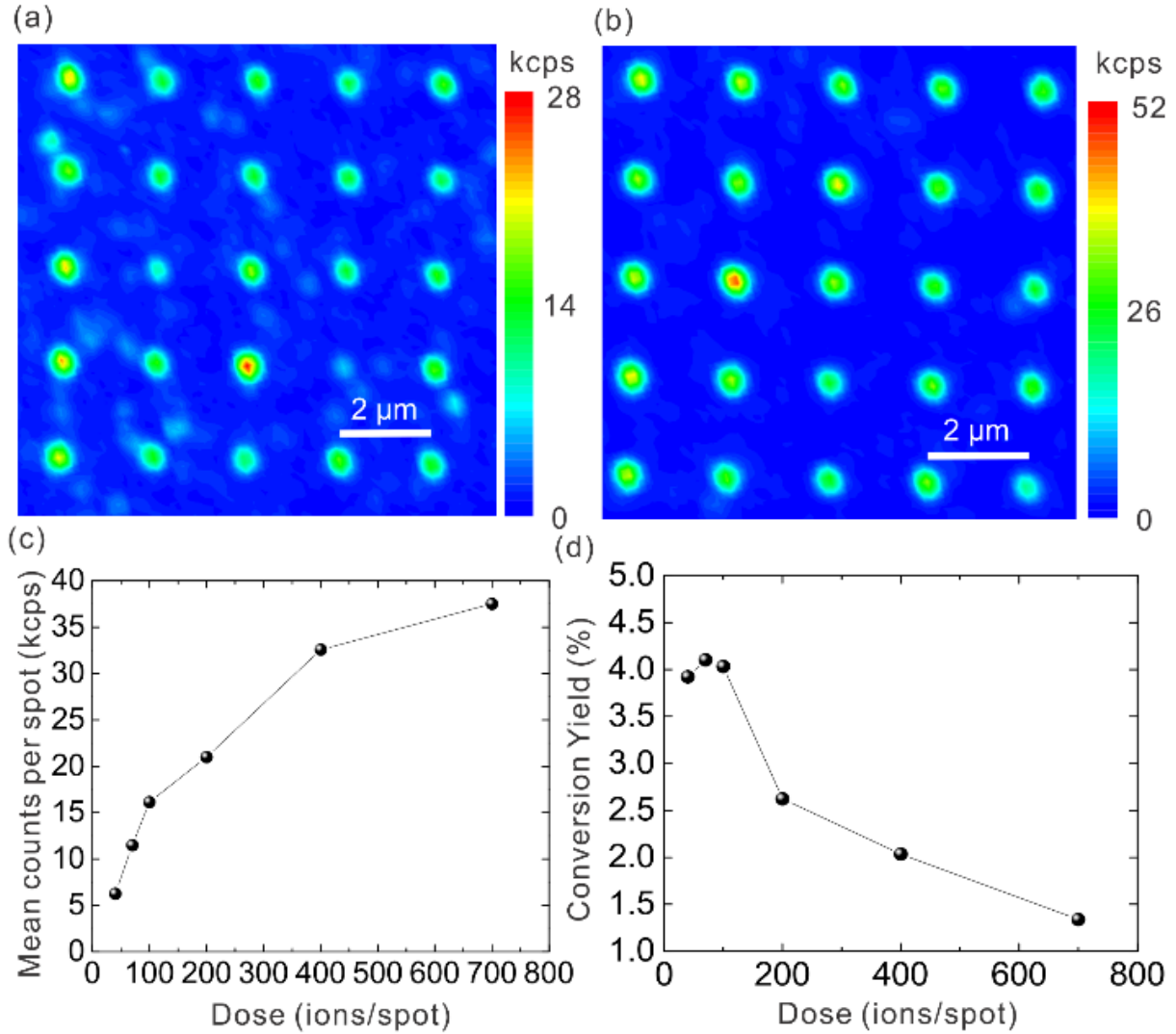

Figure 3. (a) and (b) Confocal fluorescence scans of a silicon carbide surface implanted with focused beams of Si$^{++}$ ions corresponding to an applied dose of 100 and 400 respectively. The scale bar is 2μm. (c) The mean fluorescence rates per spot as a function of Si$^{++}$ ion dose (ions/spot). (d) The conversion yield of Si$^{++}$ ions to V$_{Si}$ defects as a function of Si$^{++}$ ion dose (ions/spot).